\documentclass[aps,prl,twocolumn,groupedaddress]{revtex4-1}
\usepackage{epsfig}
\usepackage{graphicx}% Include figure files
\usepackage{dcolumn}% Align table columns on decimal point
\usepackage{bm}% bold math
\usepackage{amsthm}
\usepackage{amsfonts}
\usepackage{amscd}
\usepackage{amsmath}    % need for subequations
\usepackage{enumerate}

\newcommand{\ket}[1]{\mbox{$ | #1 \rangle $}}
\newcommand{\bra}[1]{\mbox{$ \langle #1 | $}}

\begin{document}

%\preprint{}

\title{Heralded qubit amplifiers for practical device-independent quantum key distribution}



\author{Marcos Curty$^1$ and Tobias Moroder$^2$}
%\thanks{}
\affiliation{$^1$ ETSI Telecomunicaci\'on, Department of Signal Theory and Communications, University of Vigo, 
Campus Universitario, E-36310 Vigo, Pontevedra, Spain \\
$^2$ Institut f\"ur Quantenoptik und Quanteninformation, \"Osterreichische Akademie der Wissenschaften, Technikerstra\ss{}e 21A,
A-6020 Innsbruck, Austria 
%\\
%$^3$ Center for Quantum Information and Quantum Control, Department of Physics and Department of 
%Electrical \& Computer Engineering, University of Toronto, M5S 3G4 Toronto, Ontario, Canada
}


\date{\today}

\begin{abstract}
Device-independent quantum key distribution does not need a precise quantum mechanical model of employed devices to guarantee security. Despite of its beauty, it is still a very challenging experimental task. We compare a recent proposal by Gisin \textit{et al.} [Phys. Rev. Lett. \textbf{105}, 070501 (2010)] to close the detection loophole problem with that of a simpler quantum relay based on entanglement swapping with linear optics. Our full-mode analysis for both schemes confirms that, in contrast to recent beliefs, the second scheme can indeed provide a positive key rate which is even considerably higher than that of the first alternative. 
The resulting key rates and required detection efficiencies of approx. 95\% for both schemes, however, strongly depend on the underlying security proof. 
\end{abstract}

\pacs{}
%\keywords{}

\maketitle

%\section{Introduction}\label{intro}

Despite of its often praised unconditional security, quantum cryptography also relies on some assumptions. Some of them are quite natural, such as the validity of quantum mechanics, the existence of true random number generators, or to assume that the legitimate users are well shielded from the eavesdropper. Other assumptions are more severe, like considering that the honest parties have an accurate and complete description of their physical devices. Obviously, if the functioning of the real setup differs from that considered in the mathematical model, this may become completely vulnerable to new types of attacks not covered by the security proof \cite{makarov}. 

In principle, this presumably hard-verifiable requirement of characterizing real devices can be circumvented using device-independent quantum key distribution (diQKD) \cite{diqkd1,diqkd2,pironio09}. Here, the legitimate users only need to specify a certain number of possible in- and outputs for each ``black-box'', and can prove the security of the protocol based on the violation of an appropriate Bell inequality, which certifies the presence of quantum correlations. In practice, however, diQKD is a very challenging experimental problem. Specially, it is necessary to close the detection loophole which is present in all optical tests of Bell's inequalities realized so far, even at short distances \cite{loop}. Current experimental non-locality tests use the so-called fair-sampling assumption to cope with the low efficiencies of both the quantum channel and detectors, but unfortunately this premise cannot be justified in a complete device independent scenario.

In this Letter we investigate a potential solution to bypass this detection loophole problem due to channel losses in diQKD in order to cover long distances. In particular, we compare a recent proposal by Gisin \textit{et al.} \cite{heral2} based on so-called qubit amplification, with that of a standard quantum relay which employs entanglement swapping. Contrary to recent arguments \cite{heral2,heral3}, our full-mode simulation 
for both schemes
demonstrates that the second alternative can indeed provide a positive key rate using only linear optical components. This key rate is also considerably higher than that of the first alternative. 
%Additionally, we show that photon number resolving (PNR) detectors are not 
%strictly necessary for these systems, but threshold detectors, although less efficient, are an alternative solution. 
Let us stress that our main motivation lies on experimental realizations of diQKD over long distances, rather than presenting a rigorous full security proof for such schemes in the presence of losses. For that, we employ the security analysis provided in Ref.~\cite{heral2}, which holds for specific kinds of eavesdropping attacks that are assumed, but unproven, to be optimal. For comparison reasons, we also evaluate a conservative lower bound on the secret key generation rate that can be obtained by deterministic, or random assignment of inconclusive to conclusive events \cite{pironio09}. In this last scenario, however, the employed detectors must have almost perfect efficiency in order to distribute a secret key with practical signals. These differences should further emphasize the strong performance and requirements dependence of these schemes with respect to the underlying security analysis.

\begin{figure}
\begin{center}
\includegraphics[angle=0,scale=0.53]{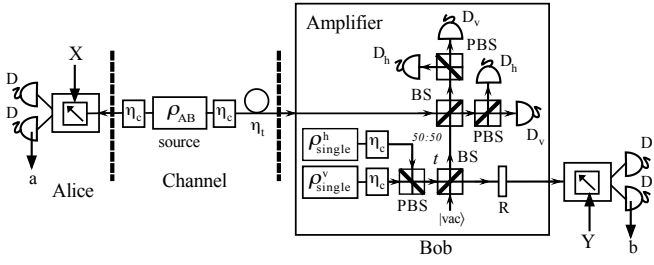}
\end{center}
\caption{Basic setup of a diQKD scheme with a quantum teleportation-based heralded qubit amplifier located on Bob's side \cite{heral2}. The entanglement source $\rho_{AB}$ is near Alice's transmitter. The parameter $\eta_c$ denotes the efficiency of optical couplers, $t$ is the transmittance of a beamsplitter (BS), PBS stands for a polarizing BS, $\rho_{\rm single}^{h}$ and $\rho_{\rm single}^{v}$ represent two single-photon sources generating horizontal ($h$) and vertical ($v$) polarized photons respectively, $R$ is a polarization rotator, and $D$ and $D_i$, with $i\in\{h,v\}$, denote photodetectors. 
The single-photon sources $\rho_{\rm single}$ can be realized, for instance, with heralded entanglement sources like spontaneous parametric down-conversion (SPDC) sources.
%Whenever Alice and Bob succeed distributing an entangled pair of photons, the amplifier outputs an heralding signal for Bob. Only when such successful event occurs, Bob chooses his measurement setting input $Y$. 
\label{figure_general}}
\end{figure}

As a starting point of our considerations let us recall the heralded qubit amplifier introduced by Gisin {\it et al.} in Ref.~\cite{heral2}, extending an earlier work by Ralph and Lund \cite{heral1}. The goal is to determine if an arriving light pulse contains precisely one photon or not, without disturbing its state of polarization. Such a scheme can be seen as a quantum-non-demolition measurement that distinguishes single-photon signals from vacuum/multiphoton pulses. 
%An empty pulse could either arise, for instance, due to losses in the quantum channel or in the source, while a multi-photon pulse typically originates from imperfections in the source, though these last events become more unlikely for longer distances. 
The basic setup is illustrated in Fig.~\ref{figure_general}.
The amplifier consists of a linear optics network, together with two single-photon sources, which are denoted in the figure as $\rho_{\rm single}^{h}$ and $\rho_{\rm single}^{v}$, emitting horizontal ($h$) and vertical ($v$) polarized photons respectively. Whenever a single-photon pulse from the channel enters the amplifier, its state of polarization is teleported  to a photon situated at its output port. If the incoming light pulse is empty or contains more than one photon, however, the teleportation process fails with high probability. By looking at the detection pattern observed in the 
photodetectors $D_i$, with $i\in\{h,v\}$, Bob can verify which of these two possible events occurred. 

Let us consider first for simplicity the scenario where all optical elements within the amplifier are lossless, and all detectors $D_i$ are noiseless, have photon number resolution and perfect detection efficiency. Moreover, let us assume that $\rho_{\rm single}^{h}$ and $\rho_{\rm single}^{v}$ emit exactly
one photon each in the correct polarization, and 
%the entangled state $\rho_{AB}$ has the form 
\begin{equation}
\rho_{AB}=(1-p)\ket{0}\bra{0}+p\ket{\frac{a_h^\dagger{}b_h^\dagger+a_v^\dagger{}b_v^\dagger}{\sqrt{2}}}
\bra{\frac{a_h^\dagger{}b_h^\dagger+a_v^\dagger{}b_v^\dagger}{\sqrt{2}}},
\end{equation}  
where $\ket{0}$ denotes the vacuum state, $a_h^\dagger{}$ and $b_h^\dagger$ ($a_v^\dagger{}$ and $b_v^\dagger$) represent the creation operators for the horizontal (vertical) polarization modes, and $0<p\leq{}1$. 
%That is, the state $\rho_{AB}$ is just a mixture of the vacuum signal and a maximally entangled state. 
In this situation, it turns out that whenever Bob's detectors $D_i$ observe two photons prepared in orthogonal polarizations, 
%({\it i.e.}, only two detectors $D_h$ and $D_v$ within the amplifier register precisely one photon each), 
the unnormalized conditional state at the input ports of Alice's and Bob's measurement devices $X$ and $Y$ (see Fig.~\ref{figure_general}) has the form (after an appropriate one-photon rotation $R$)
\begin{eqnarray}\label{condstate}
\sigma_{AB}&=&(1-p)\frac{(1-t)^2}{4}\ket{0}\bra{0}+p\frac{(1-\eta_t)(1-t)^2}{8}
\Big[\ket{a_h^\dagger}\bra{a_h^\dagger} \nonumber \\
&+&\ket{a_v^\dagger}\bra{a_v^\dagger}\Big]+p\frac{\eta_t{}t(1-t)}{4}
\ket{\frac{a_h^\dagger{}b_h^\dagger+a_v^\dagger{}b_v^\dagger}{\sqrt{2}}}
\bra{\frac{a_h^\dagger{}b_h^\dagger+a_v^\dagger{}b_v^\dagger}{\sqrt{2}}}.
\end{eqnarray}
Here, $t$ is the transmittance of a beamsplitter (BS) within the amplifier, and $\eta_t$ denotes the transmission efficiency of the quantum channel. 

By selecting a sufficiently high value for the parameter $t$, Alice and Bob can always amplify the maximally entangled component of $\sigma_{AB}$ for any transmission efficiency of the quantum channel. This technique provides them with a powerful tool to overcome the problem of transmission losses in diQKD. Any successful amplifier event acts as a kind of fair-sampling device. Since the real measurement input is only chosen afterwards, there should be no correlations between the trigger and the input choice \cite{note_t2}.
%For that, Bob simply chooses his measurement input $Y$ only when he is sure to share with Alice an entangled state due to a successful amplification event, which acts as a kind of fair-sampling device \cite{note_t2}. 
As a result, it turns out that the overall detection efficiency which is needed to close the detection loophole in diQKD can be reduced to basically that of Alice's and Bob's devices, but does not depend anymore on the loss of the quantum channel. A drawback of this technique is, however, the small success probability, $P_{\rm succ}=(1-t)[1-t-p\eta_t(1-2t)]$, of having a successful heralding signal from the amplifier for large $t$, which might strongly reduce the achievable secret key rate of the protocols. 
This imposes a trade-off on the value of the transmittance $t$. To guarantee that Alice and Bob share a high entangled state $\sigma_{AB}$ suitable for diQKD favors $t\approx{}1$, but for small transmission efficiencies this implies an almost zero success probability.
% $P_{\rm succ}$. 

\begin{figure}
\begin{center}
\includegraphics[angle=0,scale=0.53]{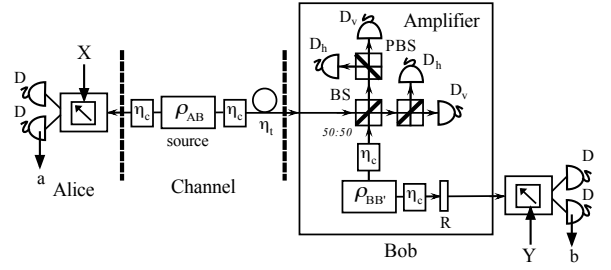}
\end{center}
\caption{Basic setup of a diQKD system with a standard quantum relay using linear optics. 
When compared to Fig.~\ref{figure_general}, now the two single-photon sources $\rho_{\rm single}^{h}$ and $\rho_{\rm single}^{v}$, together with the BS of transmittance $t$, have been replaced by just one entanglement source $\rho_{BB'}$. 
%The measurement scheme that performs a partial Bell measurement (within the amplifier) coincides with that used 
%in Fig.~\ref{figure_general}. The two single-photon sources $\rho_{\rm single}^{h}$ and $\rho_{\rm single}^{v}$, together with the BS of transmittance $t$, however, have now been replaced by just one entanglement source $\rho_{BB'}$. 
\label{figure_general2}}
\end{figure}

A more direct approach to implement a heralded qubit amplifier is to use a standard quantum relay with linear optics. The basic setup is illustrated in Fig.~\ref{figure_general2}. 
The working principle of this scheme is essentially the same as that of a teleportation-based amplifier. The only difference between both solutions relies on Bob's mechanism to generate an entangled state in the amplifier for teleportation. While in Ref.~\cite{heral2} Bob mixes two-photon pulses with a vacuum signal at a BS whose transmittance is optimized, in a quantum relay architecture he directly uses an entanglement photon source, which might be easier to realize experimentally. Intuitively speaking, one could expect that a linear optics quantum relay might be valuable for long distance diQKD {\it only} when Alice and Bob have a high-quality entanglement source at their disposal. Otherwise, the conditional signals $\sigma_{AB}$ shared by the legitimate users (after a successful amplifier event) might be poorly entangled. That is the case, for instance, when $\rho_{AB}$ and $\rho_{BB'}$ (see Fig.~\ref{figure_general2}) are generated with spontaneous parametric down-conversion (SPDC) sources. We will show that this intuition is wrong, and a quantum relay can indeed be used to achieve considerable higher secret key rates  than those obtained with the teleportation-based amplifier of Fig.~\ref{figure_general}, even with practical signals. 

Let us begin again by considering a simplified scenario where all detectors $D_i$ are noiseless, photon number resolving and perfectly efficient. Moreover, we assume that 
$\eta_t=1$ and the states $\rho_{AB}=\rho_{BB'}$ have the form
\begin{eqnarray}\label{tuesn}
\rho_{AB}&=&p_0\ket{0}\bra{0}+p_1\ket{\frac{a_h^\dagger{}b_h^\dagger+a_v^\dagger{}b_v^\dagger}{\sqrt{2}}}
\bra{\frac{a_h^\dagger{}b_h^\dagger+a_v^\dagger{}b_v^\dagger}{\sqrt{2}}}
\nonumber \\
&+&p_2
\ket{\frac{(a_h^\dagger{}b_h^\dagger+a_v^\dagger{}b_v^\dagger)^2}{2\sqrt{3}}}
\bra{\frac{(a_h^\dagger{}b_h^\dagger+a_v^\dagger{}b_v^\dagger)^2}{2\sqrt{3}}},
\end{eqnarray}
with $p_0+p_1+p_2=1$.
In this situation, it can be shown that whenever Bob's detectors $D_i$ observe precisely two photons prepared in orthogonal polarizations, the unnormalized conditional quantum state shared with Alice is given by
\begin{eqnarray}\label{lun}
\sigma_{AB}&=&\frac{p_0p_2}{3}\Big[\ket{a_h^\dagger{}a_v^\dagger}\bra{a_h^\dagger{}a_v^\dagger}
+\ket{b_h^\dagger{}b_v^\dagger}\bra{b_h^\dagger{}b_v^\dagger}\Big]
\nonumber \\
&+&
\frac{p_1^2}{2}
\ket{\frac{a_h^\dagger{}b_h^\dagger+a_v^\dagger{}b_v^\dagger}{\sqrt{2}}}
\bra{\frac{a_h^\dagger{}b_h^\dagger+a_v^\dagger{}b_v^\dagger}{\sqrt{2}}}.
\end{eqnarray}

In contrast to Eq.~(\ref{condstate}), there is no parameter now that Alice and Bob could tune to amplify further the maximally entangled component of $\sigma_{AB}$. Actually, depending on the probability distribution $p_n$ of the entanglement sources $\rho_{AB}$ and $\rho_{BB'}$, the fidelity of the outgoing state $\sigma_{AB}$ with respect to a maximally entangled state could be very low. This occurs, for example, when Bob uses SPDC sources with a photon number distribution given by $p_n=(n+1)\lambda^n/(1+\lambda)^{n+2}$, where $\lambda$ denotes a parameter related to the pump amplitude of the laser. For small values of $\lambda$, this fidelity is roughly equal to %$F=\bra{(a_h^\dagger{}b_h^\dagger+a_v^\dagger{}b_v^\dagger)
%/\sqrt{2}}\sigma_{AB}\ket{(a_h^\dagger{}b_h^\dagger+a_v^\dagger{}b_v^\dagger)/\sqrt{2}}
%\approx{}1/2$. 
$1/2$. The scenario changes if Alice and Bob post-select only those detection events where both of them see precisely one photon in their measurement devices $X$ and $Y$. However, such strategy seems to open the detection loophole. Note that the probability $\mu_{cc}$ that Alice and Bob obtain a conclusive result ({\it i.e.}, both of them observe exactly one photon each in their measurement apparatuses) is also about $1/2$ for small values of $\lambda$. This result is far below the typical detection efficiency of $82.8\%$ that is required to violate the Clauser-Horne-Shimony-Holt (CHSH) inequality \cite{clauser} that prevents eavesdropping exploiting the detection loophole. This argumentation seems to render the conditional signal states $\sigma_{AB}$ given by Eq.~(\ref{lun}) unsuitable for diQKD \cite{heral2,heral3}.

The key point here, however, is simple, but counterintuitive: the detection efficiency limit of $82.8\%$ does not apply to the correlations observed when measuring the signals $\sigma_{AB}$. This can be seen with a simple example. Suppose, for instance, that Alice and Bob employ a post-processing strategy where inconclusive outcomes are assigned to conclusive ``$+1$'' outcomes in a deterministic fashion \cite{pironio09}. In this situation, the CHSH quantity becomes 
\begin{equation}
S=\mu_{cc}S_{cc}+2(1-\mu_{cc}), %=1+\sqrt{2}>2,
\end{equation}
where the parameter $S_{cc}$ denotes the CHSH value computed only on the set of conclusive results obtained before applying the post-processing step. For small $\lambda$, we find that $S$ is roughly given by $S=1+\sqrt{2}>2$.  That is, Alice and Bob can indeed detect the presence of non-local correlations in the signals $\sigma_{AB}$. This result mainly arises because $\sigma_{AB}$ provides Alice and Bob with an atypical detection pattern: either both of them obtain a conclusive or an inconclusive outcome. But no conclusive-inconclusive or 
inconclusive-conclusive outcomes are observed, as typically present for ``local detection losses''. This argument actually holds for any value of $p_1>0$ in Eq.~(\ref{tuesn}). 
When the detection efficiency of Alice's and Bob's detectors is not perfect (but high enough), it turns out that the probability to observe conclusive-inconclusive or 
inconclusive-conclusive results is very low, and they can still violate the CHSH inequality and distribute a secret key. 
%but observations like conclusive-inconclusive, as present for ``local detection losses'', are rather rare provided that the detection efficiency if sufficiently high. Let us point out that this argument holds for any non-vanishing $p_1>0$ in the state of Eq.~\ref{lun}.

\begin{figure}
\begin{center}
\includegraphics[angle=0,scale=0.35]{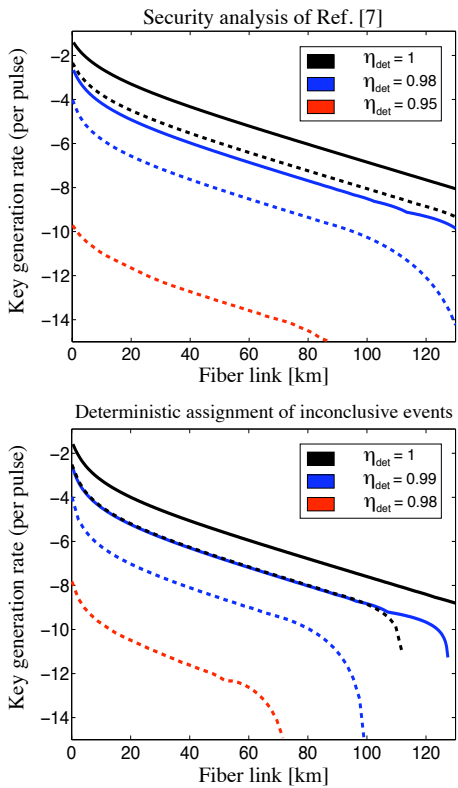}
\end{center}
\caption{Lower bound on the secret key rate in logarithmic scale (base $10$) vs. distance for a diQKD setup using the amplifier illustrated in Fig.~\ref{figure_general} (dashed line), and the quantum relay with linear optics shown in Fig.~\ref{figure_general2} (solid line). The upper figure corresponds to the security analysis provided in Refs.~\cite{heral2, note3}. The lower one represents the situation where the legitimate users assign inconclusive to conclusive results deterministically \cite{pironio09}. In this last case, the minimum detection efficiency of Alice's and Bob's detectors is around $98\%$ (for the teleportation-based amplifier) and 
$99\%$ (for the quantum relay). In the simulations we assume $\eta_c=\eta_{\rm det}$ and the loss coefficient of the optical fiber  $\alpha=0.2$ dB/km. 
\label{figure_original}}
\end{figure}

To evaluate the performance of both setups in a more realistic situation, we perform a full-mode analysis for the case where Alice and Bob use both entangled and heralded single-photon sources based on SPDC, together with inefficient photon number resolving detectors. For simplicity, however, we do not consider any misalignment effect in the quantum channel or in Alice's and Bob's detection apparatuses (in view that photon loss is the dominant error mechanism), and we also neglect the effect of dark counts in the photodetectors (which typically results in a cut-off distance where the dark-count free key generation rate and the overall dark count probability are roughly equal). To compute a lower bound on the secret key rate we employ the diQKD protocol based on the violation of the CHSH inequality analyzed in Refs.~\cite{diqkd2,pironio09,diqkd3}, and we evaluate two different secret key rate formulas. The first follows the security analysis presented in Ref.~\cite{heral2} and holds for particular eavesdropping attacks that are assumed to be optimal \cite{note3}. 
%The first one was derived in Ref.~\cite{heral2} and holds for particular eavesdropping attacks that are assumed to be optimal. 
The second one corresponds to the conservative situation where the legitimate users assign inconclusive to conclusive events in a deterministic fashion \cite{pironio09}. For the cases studied, this last strategy seems to perform better than that based on a purely random assignment of inconclusive to conclusive outcomes \cite{random}. The results are illustrated in Fig.~\ref{figure_original} for a few values of the coupling efficiency $\eta_c$ and the detection efficiency $\eta_{\rm det}$ of Alice's and Bob's detectors. In these simulations we consider only the contribution till four-photon pairs emitted by the SPDC sources. That is, we take $p_n=0$ for all $n\geq{}5$, which is reasonable when $\lambda$ satisfies $\lambda\ll{}1$. For a given distance, we optimize the transmittance $t$ and the intensities of the different lasers numerically to maximize the resulting key rate. When $\eta_c$ and $\eta_{\rm det}$ are high enough, Fig.~\ref{figure_original} shows that the use of a quantum relay can provide significantly higher key rates than a teleportation-based amplifier, while this last alternative can tolerate slightly lower detection efficiencies (around $95\%$) than the former one (around $96\%$), though the achievable key rates in this regime are already quite low. The improved performance of the quantum relay in comparison with the amplifier scheme seems to rest mainly on the quality of the single-photon sources $\rho_{\rm single}^{h}$ and $\rho_{\rm single}^{v}$. 
Only if one 
considers the ideal scenario where these sources are perfect and on-demand, then the second scheme can deliver similar key rates as those of a quantum-relay with practical SPDC sources. However, when Bob uses heralded single-photon sources based on SPDC instead (see Fig.~\ref{figure_original}), the small probability to find one photon in the idler mode of both sources at the same time strongly reduces the resulting key rate 
of a teleportation-based amplifier.
Although less efficient, the use of threshold detectors might be also an alternative to photon number resolving detectors. However, the analysis of this scenario is more involved since the probability that Alice and Bob obtain conclusive/inconclusive events depends on the basis choice. Details of this analysis will be presented somewhere else.
 
To conclude, we have performed a full-mode analysis of two potential solutions to circumvent the 
problem of transmission losses in diQKD using only linear optical components. Contrary 
to recent findings, we have demonstrated that a standard quantum relay is indeed 
an alternative and can outperform  
a teleportation-based amplifier. Still, a main technological challenge 
here
is to develop photodetectors with nearly perfect 
detection efficiency and negligible noise. Recent results in this field give
reasons to be optimistic \cite{det}. 

The authors specially thank H.-K. Lo for bringing the subject of heralded 
qubit amplification to our attention. 
%His enlightening discussions started our investigation. 
We also thank 
S. Pironio for explaining to us the security analysis presented in Ref.~\cite{heral2} and 
for 
stimulating discussions on this topic, together with 
N. Sangouard and N. L\"utkenhaus. We are indebt to J. M. Taboada and
F. Obelleiro 
for their support in the use of a cluster 
to run the simulations.  
This work was supported by Xunta de Galicia, Spain (Grant 
No. INCITE08PXIB322257PR), and by the FWF (START Prize and SFB FOQUS).


\begin{thebibliography}{99}
\bibitem{note_t2}
If the measurement setting is chosen in advance, the trigger outcome could again depend on the input choice. That is, a real experiment 
should not post-select those events where the trigger has succeeded.
\end{thebibliography}
\bibliographystyle{apsrev}

\end{document}